\documentclass[onecolumn,showpacs,10pt]{revtex4}

\usepackage{graphicx}
\usepackage{dcolumn}
\usepackage{bm}

\input epsf

\begin{document}

\title{\Large Dynamical System Analysis for Anisotropic Universe in Brans-Dicke Theory}

\author{\bf Jhumpa Bhadra$^1$\footnote{bhadra.jhumpa@gmail.com},
Shuvendu Chakraborty$^2$\footnote{shuvendu.chakraborty@gmail.com}
and Ujjal Debnath$^1$\footnote{ujjaldebnath@yahoo.com ,
ujjal@iucaa.ernet.in}}

\affiliation{$^1$ Department of Mathematics,
Bengal Engineering and Science University, Shibpur, Howrah-711
103, India.\\ $^2$Department of Mathematics, Seacom Engineering
College, Howrah-711 302, India.}

\date{\today}

\begin{abstract}
In this work, we have studied the Brans-Dicke (BD) cosmology in
anisotropic models. We present three dimensional dynamical system
describing the evolution of anisotropic models containing perfect
fluid and BD scalar field with self-interacting potential. The
relevant equations have been transformed into the dynamical
system. The critical points and the corresponding eigen values
have been found in radiation, dust, dark energy, $\Lambda$CDM and
phantom phases of the universe. The natures and the stability
around the critical points have also been investigated.
\end{abstract}


\maketitle

\section{\normalsize\bf{Introduction}}

A recent renewal of interest in Brans-Dicke (BD) theory [1] can be
traced to the discovery by La and Steinhardt [2] that the use of
BD theory in place of general relativity can ameliorate the exit
problem of inflationary cosmology. This is possible because the
interaction of the BD scalar field with the metric slows the
expansion from exponential to power-law. The BD theory contains
only one dimensionless parameter, $\omega$ and the effective
gravitational constant is inversely proportional to the scalar
field $\phi$. BD theory has been proved to be very effective
regarding the recent study of cosmic acceleration [3]. This theory
yields the correct Newtonian weak-field limit, but solar system
measurements of post-Newtonian corrections require $\omega>500$
[4]. In the limit $\omega \rightarrow \infty$, the field $\phi$
becomes a constant [5] and we recover Einstein gravity. This
theory has very effectively solved the problems of inflation and
the early and the late time behaviour of the Universe. N. Banerjee
and D. Pavon [3] have shown that BD scalar tensor theory can
potentially solve the quintessence problem. The generalized BD
theory [6] is an extension of the original BD theory with a time
dependent coupling function $\omega$. In Generalized BD theory,
the BD parameter $\omega$ is a function of the scalar field
$\phi$. This has led to more general scalar-tensor gravity [7]
being considered with a self-interacting potential [8, 9].
Modified BD theory with a self-interacting potential have also
been introduced in this regard. Bertolami and Martins [10] have
used this theory to present an accelerated Universe for spatially
flat model. They have obtained the solution for accelerated
expansion with a potential ${\phi}^{2}$ and large $|\omega|$,
although they have not considered the positive energy conditions
for the matter and
scalar field.\\

Dynamical system theory [11] has been applied with great success
in cosmology and astrophysics within the context of general
relativity. This theory are used to describe the behaviour of
complex dynamical systems usually by constructing differential
equations. This theory deals with a long term qualitative
behaviour of the formed differential equations. It does not
concentrate to find the precise solutions of the system but
provide answers like- whether the system is stable for long time
and whether the stability depend on the initial conditions.
Besides the other scientific fields this theory is now become
widely useful in the research of cosmology. Its range of
applicability has been enlarged by considering alternate theories
of gravity, such as those which do not obey the principle of
minimal coupling [12], Kaluza-Klein [13] and scalar-tensor
theories [14]. There are several works on qualitative analysis in
dynamical system of FRW cosmology in BD gravity [9, 15]. There are
some literature where the authors have studied dynamical evolution
of some models where the universe is filled with dark energy on
an interaction has occurred between dark matter or barotropic
matter and some dark energies like Generalized Chaplygin Gas [16],
New Generalized Chaplygin Gas[17], ELKO non standard spinor dark
energy [18], DBI essence [19], non- minimally coupled scalar field
[20], k-essence [21] etc. There are some other approach of
dynamical system analysis to diffrent model the universe on the
frame work of loop quantum gravity and non linear electro dynamics
[22, 23]. Some authors [9, 24] have shown that the field equations
for FRW cosmological models could be reduced to a two dimensional
dynamical system for any reasonable perfect fluid matter source in
BD theory and in the presence of scalar potential, the
cosmological dynamical system turns out to be generally three
dimensional, apart from radiation dominated universes in which the
system once more reduces to two dimensions.\\

Here we have considered an anisotropic model [25, 26] containing
perfect fluid and BD scalar field with self-interacting potential.
The starting point of our work is to reduce the field equations to
a three dimensional autonomous dynamical system. From the
knowledge of mathematical features of the system, we have found
the critical points and the exact solutions for the system of
equations. Some of the solutions are valid for fluids
satisfying the equation of state and for particular values of $\gamma$.
The stability around the critical points have been investigated.\\

\section{\normalsize\bf{Basic Equations}}

The self-interacting Brans-Dicke (BD) theory is described by the
action: (choosing $8\pi G_{0}=c=1$)

\begin{equation}
S=\int d^{4} x \sqrt{-g}\left[\phi R- \frac{\omega}{\phi}
{\phi}^{,\alpha} {\phi,}_{\alpha}-V(\phi)+ {\cal L}_{m}\right]
\end{equation}

where $V(\phi)$ is the self-interacting potential for the BD
scalar field $\phi$ and $\omega$ is the BD parameter. The matter
content of the Universe is composed of perfect fluid,
\begin{equation}
T_{\mu \nu}=(\rho+p)u_{\mu} u_{\nu}+p~g_{\mu \nu}
\end{equation}

where $u_{\mu}~u^{\nu}=-1$ and $\rho,~p$ are respectively energy
density and isotropic pressure.\\

From the Lagrangian density $(1)$ we obtain the field equations
\begin{equation}
G_{\mu \nu}=\frac{\omega}{{\phi}^{2}}\left[\phi  _{ , \mu} \phi
_{, \nu} - \frac{1}{2}g_{\mu \nu} \phi _{, \alpha} \phi ^{ ,
\alpha} \right] +\frac{1}{\phi}\left[\phi  _{, \mu ; \nu} -g_{\mu
\nu}~ ^{\fbox{}}~ \phi \right]-\frac{V(\phi)}{2 \phi} g_{\mu
\nu}+\frac{1}{\phi}T_{\mu \nu}
\end{equation}
and
\begin{equation}
^{\fbox{}}~\phi=\frac{1}{3+2\omega}T+\frac{1}{3+2\omega
}\left[\phi
 \frac{dV(\phi)}{d\phi}-2V(\phi)\right]
\end{equation}

where $T=T_{\mu \nu}g^{\mu \nu}$.\\

We  consider homogeneous and anisotropic  space-time  model
described by the line  element

\begin{equation}
ds^{2}=-dt^{2}+a^{2}dx^{2}+b^{2}d\Omega_{k}^{2}
\end{equation}

where  $a$  and  $b$  are  functions  of  time  $t$ alone : we
note that

\begin{eqnarray}d\Omega_{k}^{2}= \left\{\begin{array}{lll}
dy^{2}+dz^{2}, ~~~~~~~~~~~~ \text{when} ~~~k=0 ~~~~ ( \text{Bianchi ~I ~model})\\
d\theta^{2}+sin^{2}\theta d\phi^{2}, ~~~~~ \text{when} ~~~k=+1~~
( \text{Kantowaski-Sachs~ model})\\
d\theta^{2}+sinh^{2}\theta d\phi^{2}, ~~~ \text{when} ~~~k=-1 ~~(
\text{Bianchi~ III~ model})\nonumber
\end{array}\right.
\end{eqnarray}

Here  $k$  is  the  curvature  index  of  the  corresponding
2-space, so  that  the  above  three  types  are  described  by
Thorne [19]  as  flat, closed  and  open respectively.\\

Now, in  BD  theory, the Einstein's  field  equations  for  the
above space-time  symmetry  are

\begin{equation}
\frac{\ddot{a}}{a}+2\frac{\ddot{b}}{b}=-\frac{1}{(3+2\omega)\phi}\left[
(2+\omega)\rho+3(1+\omega)p\right]-\omega\frac{\dot{\phi}^{2}}
{\phi^{2}}-\frac{\ddot{\phi}}{\phi}+\frac{V(\phi)}{2\phi}
\end{equation}

\begin{equation}
\frac{\dot{b}^{2}}{b^{2}}
+2\frac{\dot{a}}{a}\frac{\dot{b}}{b}=\frac{\rho}{\phi}-\frac{k}{b^{2}}-\left(\frac{\dot{a}}{a}+2\frac{\dot{b}}{b}
\right)\frac{\dot{\phi}}{\phi}+\frac{\omega}{2}\frac{\dot{\phi}^{2}}{\phi^{2}}
+\frac{V(\phi)}{2\phi}
\end{equation}

and the wave equation for the BD scalar field $\phi$ is given by

\begin{equation}
\ddot{\phi}+\left(\frac{\dot{a}}{a}+2\frac{\dot{b}}{b}\right)\dot{\phi}=\frac{1}{3+2\omega}\left[(\rho-3p)-\frac{\phi
dV(\phi)}{d\phi}+2V(\phi)\right]
\end{equation}

The energy conservation equation is

\begin{equation}
\dot{\rho}+\left(\frac{\dot{a}}{a}+2\frac{\dot{b}}{b}\right)(\rho+p)=0
\end{equation}

Here we consider the Universe to be filled with barotropic fluid
with EOS

\begin{equation}
p=(\gamma-1)\rho~,~~~0\le\gamma\le 2
\end{equation}

The conservation equation $(9)$ yields the solution for $\rho$ as

\begin{equation}
\rho=\rho_{0}\left(ab^{2}\right)^{-\gamma}
\end{equation}

where, $\rho_{0}$ is an integration constant.

\section{\normalsize\bf{Dynamical System Analysis}}

In this section, we shall define some variables such that the
above equations can be transformed to first order differential
equations and next we investigate the dynamical system analysis.
Let us consider, $V(\phi)=V_{0}\phi^{n+1}$ and define the
following variables

\begin{equation}
X=\frac{\phi'}{\phi}
\end{equation}

\begin{equation}
Y=\frac{1}{3}\left(\frac{a'}{a}+2\frac{b'}{b}\right)+\frac{\phi'}{2\phi}
\end{equation}

where,~
$'\equiv\frac{d}{d\eta}\equiv\left(ab^{2}\right)^{1/3}\frac{d}{dt}
$~, $\eta$ is the conformal time. If $X>0$ then $Y$ must be positive. If $X<0$ then $Y$ may or may not be positive.\\

Using the transformations (12) and (13), the equations (6) and (7)
become the following two transformed equations (for $k=0$):

\begin{equation}
X'+2XY=\left(ab^{2}\right)^{2/3}\left[\frac{(4-3\gamma)}{(3+2\omega)}\frac{\rho}{\phi}-\frac{n-1}{3+2\omega}V_{0}\phi^{n}\right]
\end{equation}

\begin{equation}
Y'+2Y^{2}=\left(ab^{2}\right)^{2/3}\left[\frac{\{5+3\omega-(2+\omega)\gamma\}}{3+2\omega}\frac{\rho}{\phi}+\frac{(10+6\omega-n)}{6(3+2\omega)}V_{0}\phi^{n}\right]
\end{equation}

Also for simplicity, let us choose $n=\frac{2}{3\gamma-2}$ and
defining another transformation

\begin{equation}
Z=V_{0}\phi\left(ab^{2}\right)^{\gamma-2/3}~,
\end{equation}

(which is always positive) the above equations reduce to the
system of non-linear first order differential equations

\begin{equation}
X'+2XY=\frac{4-3\gamma}{3+2\omega}\left[\frac{V_{0}}{Z}-\frac{V_0}{3\gamma-2}\left(\frac{Z}{V_0}\right)^{\frac{2}{3\gamma-2}}\right]
\end{equation}

\begin{equation}
Y'+2Y^{2}=\frac{\{5+3\omega-(2+\omega)\gamma\}}{3+2\omega}\frac{V_{0}}{Z}+\frac{\{(3\gamma-2)(5+3\omega)-1\}V_{0}}{3(3+2\omega)(3\gamma-2)}
\left(\frac{Z}{V_{0}}\right)^{\frac{2}{3\gamma-2}}
\end{equation}

\begin{equation}
Z'=(3\gamma-2)ZY-\frac{3\gamma-4}{2}ZX
\end{equation}
\\

Now in special cases, we consider radiation, dust, dark energy,
$\Lambda$CDM and phantom models and find the critical points, eigen values and its natures separately.\\

{\bf{Case (a):}}  Dark Energy: For $\gamma=1/3$, the system of
equations (20) becomes

\begin{eqnarray}
X'=-2XY+\frac{3V_0}{Z(3+2\omega)}+\frac{3V_0^3}{Z^2 (3+2\omega)}\nonumber\\
Y'=-2Y^2+\frac{(13+8\omega)V_0}{3Z(3+2\omega)}+\frac{(2+3\omega)V_0^3}{3Z^2(3+2\omega)}\nonumber\\
Z'=\frac{3}{2}X Z-Y Z
\end{eqnarray}

There exists two critical point namely $\left(X_i^{DE},Y_i^{DE},Z_i^{DE}\right)|_{i=1,2}=\left(\mp\frac{\sqrt{2}\beta}{3},\mp\frac{\beta}{\sqrt{2}}, \frac{3(5-2\omega)V_0^2}{16\omega-1}\right)$ where $\beta=\frac{\sqrt{(16\omega-1)(7+5\omega)V_0}}{\sqrt{(5-2\omega)\{15-4(\omega-1)\omega\}}}$ and $\frac{1}{16}<\omega<\frac{5}{2}$.

Characteristic Equation around the critical point $\left(X_i^{DE},Y_i^{DE},Z_i^{DE}\right)|_{i=1,2}$ are given by the equation
\begin{equation}
\xi ^3+A_{1i} \xi^2+A_{2i} \xi+A_{3i}=0 ,~~~~~~~ i=1,2
\end{equation}

where
\begin{eqnarray}
A_{1i}=\pm\frac{\sqrt{2(16\omega-1)(7+5\omega)}}{6(5-2\omega)^{\frac{5}{2}} (3+2\omega)V_0^3\sqrt{15-4(\omega-1)\omega}}~~~~~~~~ \nonumber\\
A_{2i}=\mp\frac{(16\omega-1)(32\omega^2-442\omega-449)}{324 (5-2\omega)^4 (3+2\omega)^2 V_0^{\frac{7}{2}} }~~~~~~~~~i=1,2 \nonumber\\
A_{3i}=\mp\frac{(16\omega-1)^{\frac{5}{2}}(7+5\omega)}{81 (5-2\omega)^{\frac{7}{2}}(3+2\omega)^2 V_0^4\sqrt{2\{15-4(\omega-1)\omega\}}}
\end{eqnarray}

The condition for the existence of stable attractor solution for the dynamical system (20) is
$A_{1i}>0$, $A_{2i}>0$ and $A_{1i}A_{2i}-A_{3i}>0$ , for $i=1,2$.

In our model, the dynamical system (20) is not stable around
$\left(X_1^{DE},Y_1^{DE},Z_1^{DE}\right)$ but stable attractor in
the late time universe around the critical point
$\left(X_2^{DE},Y_2^{DE},Z_2^{DE}\right)$. Corresponding condition
becomes as follows:

\begin{eqnarray}
\frac{(16\omega-1)^\frac{3}{2}(7+5\omega)^\frac{1}{2}\left\{499+442\omega-32\omega^2-12(5-2\omega)^3(3+2\omega)(16\omega-1)V_0^\frac{5}{2}\right\}}
{972\sqrt{2}(5-2\omega)^\frac{13}{2}(3+2\omega)^3V_0^\frac{13}{2}\{15-4(\omega-1)\omega\}^\frac{1}{2}}>0
\end{eqnarray} \\

In particular, for our choice
$\gamma=\frac{1}{3},~~\omega=0.5,~~V_0=0.5$, critical point is
$(0.679563, 1.01934 , 0.428571)$ and the corresponding eigenvalues
are $(-4.07738, -1.85528, -0.183411)$, which shows a stable
attractor. Another critical point is $(-0.679563, -1.01934 ,
0.428571)$, so the system is unstable around that fixed point as
the eigenvalues are $(4.07738, 1.85528, 0.183411)$.
The phase space diagram of parameters $X(\eta), Y(\eta), Z(\eta)$ and their progressions have been drawn in figures 1 and 2 respectively. \\

\vspace{1in}
\begin{figure}[!h]

\epsfxsize = 3.0 in \epsfysize = 2 in
\epsfbox{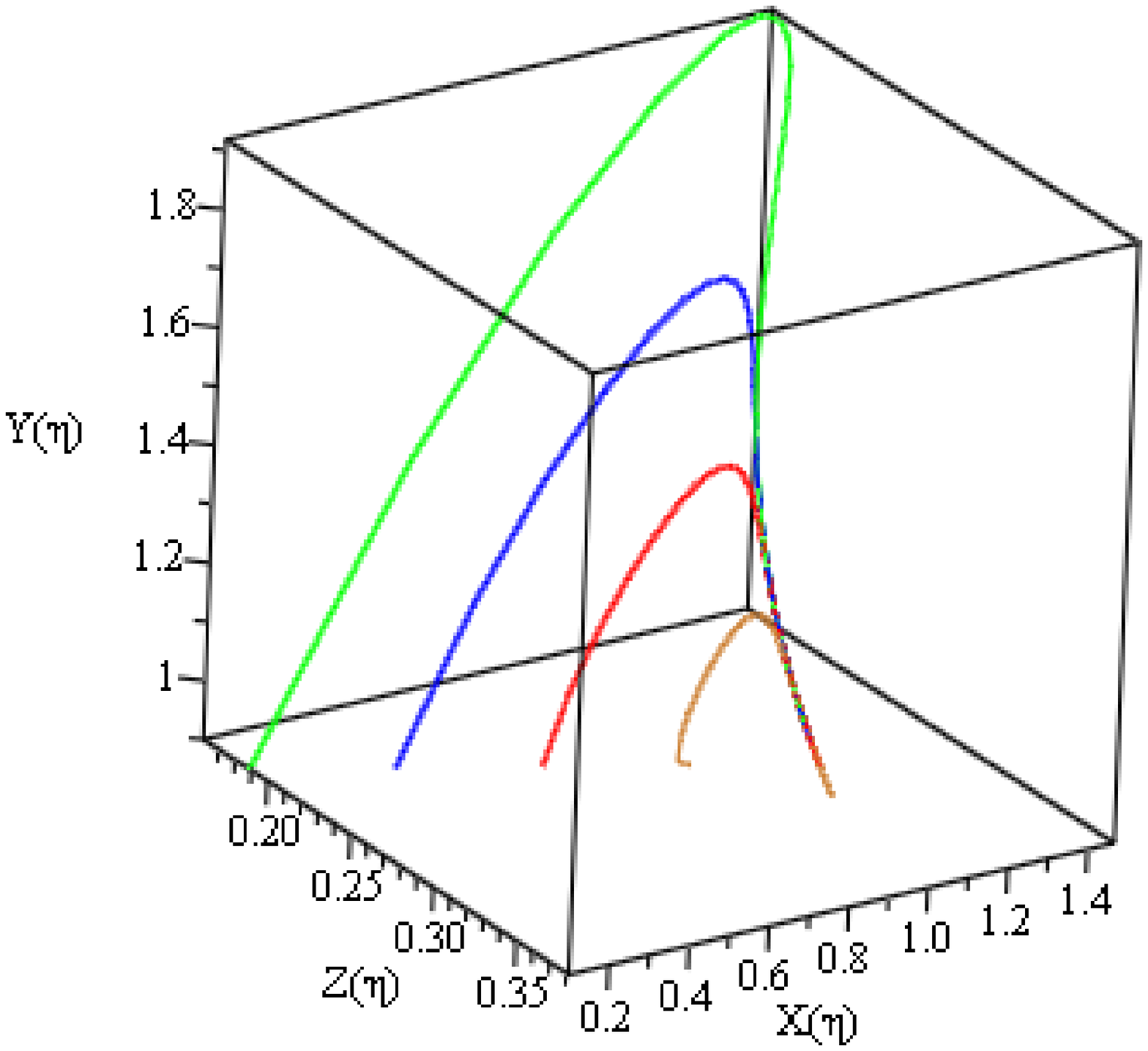}~~~\epsfxsize = 2.5 in \epsfysize = 1.7 in
\epsfbox{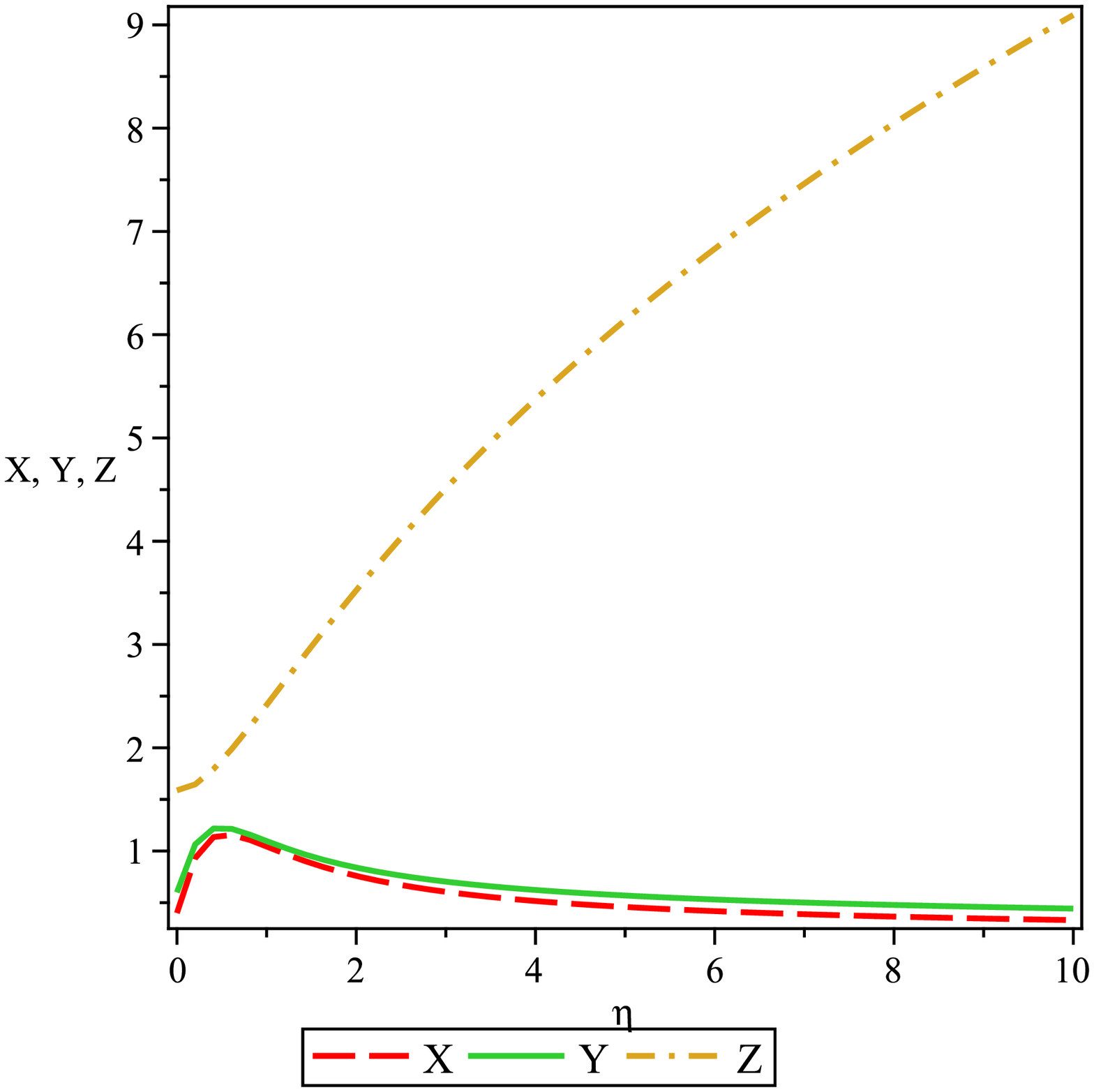}\\
~~FIG.1~~~~~~~~~~~~~~~~~~~~~~~~~~~~~~~~~~~~~~~~~~~~~~~~~~~~~~~~~~~~~~~~~~~~FIG.2\\
\caption{The phase space diagram of parameters $X(\eta), Y(\eta), Z(\eta)$ for $\gamma=\frac{1}{3}, \omega=0.5,~~V_0=0.5$.
The initial conditions chosen are $X(0) =
0.1$, $Y(0) = 0.9$, $Z(0)=0.3 V_0^{\frac{2}{3}}$ (green); $X(0) = 0.2$, $Y(0) = 1$, $Z(0)=0.4 V_0^{\frac{2}{3}}$ (blue);
$X(0) = 0.3$, $Y(0) = 1.1$, $Z(0)=0.5 V_0^{\frac{2}{3}}$ (red); $X(0) = 0.4$, $Y(0) = 1.2$, $Z(0)=0.6 V_0^{\frac{2}{3}}$(brown). }

\caption{The progression of $X(\eta), Y(\eta), Z(\eta)$ for $\gamma=\frac{1}{3},\omega=0.2,~~V_0=2$ and initial condition is $X(0) =
0.4$, $Y(0) = 0.6$, $Z(0)=V_0^{\frac{2}{3}}$ . }
\end{figure}

{\bf{Case (b):}} Phantom: For $\gamma=-1/2$, the dynamical
equation (20) becomes:

\begin{eqnarray}
X'=-2XY+\frac{11 V_0}{2(3+2\omega)Z} +\frac{11 V_0}{7(3+2\omega)Z}\left(\frac{Z}{V_0}\right)^{-\frac{1}{7}}\nonumber\\\nonumber\\
Y'=-2Y^2+\frac{(12+7\omega) V_0}{2(3+2\omega)Z} +\frac{(37+21\omega) V_0}{21(3+2\omega)Z}\left(\frac{Z}{V_0}\right)^{-\frac{1}{7}}\nonumber\\\nonumber\\
Z'=-\frac{7}{2}YZ+\frac{11}{4}ZX~~~~~~~~~~~~~~~~~~~~
\end{eqnarray}

For $\omega=-\frac{1}{2}$, there exists critical point $(X^{P},
Y^{P}, Z^{P})$ (superscript $``P"$ stands for phantom) for
positive potential function $V_0$.

In particular, $\gamma=-\frac{1}{2},~~\omega=-0.5,~~V_0=4$,
critical point is $(2.8606, 2.24761, 1.497)$ and the corresponding
eigenvalues are $(-8.99046, -4.46316, -0.0320722)$, which shows a
stable attractor. Another critical point is $(-2.8606, -2.24761,
1.497)$ system is unstable around that fixed point as the
eigenvalues are $(8.99046, 4.46316, 0.0320722)$. The phase space
diagram of parameters $X(\eta), Y(\eta), Z(\eta)$ and their
progressions
have been drawn in figures 3 and 4 respectively. \\

\vspace{1in}
\begin{figure}[!h]

\epsfxsize = 3.3 in \epsfysize = 3 in
\epsfbox{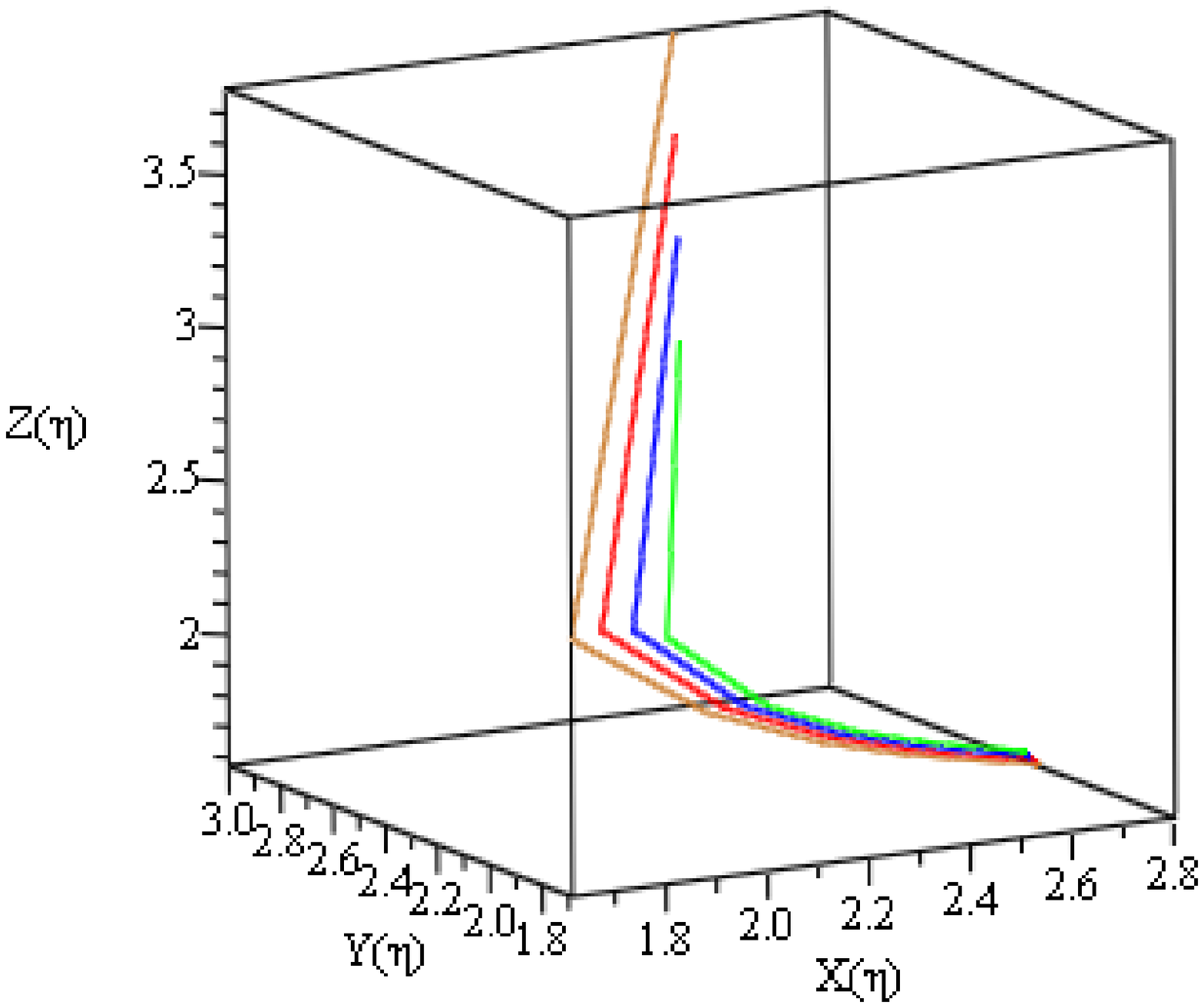}~~~\epsfxsize = 2.5 in \epsfysize = 1.7 in
\epsfbox{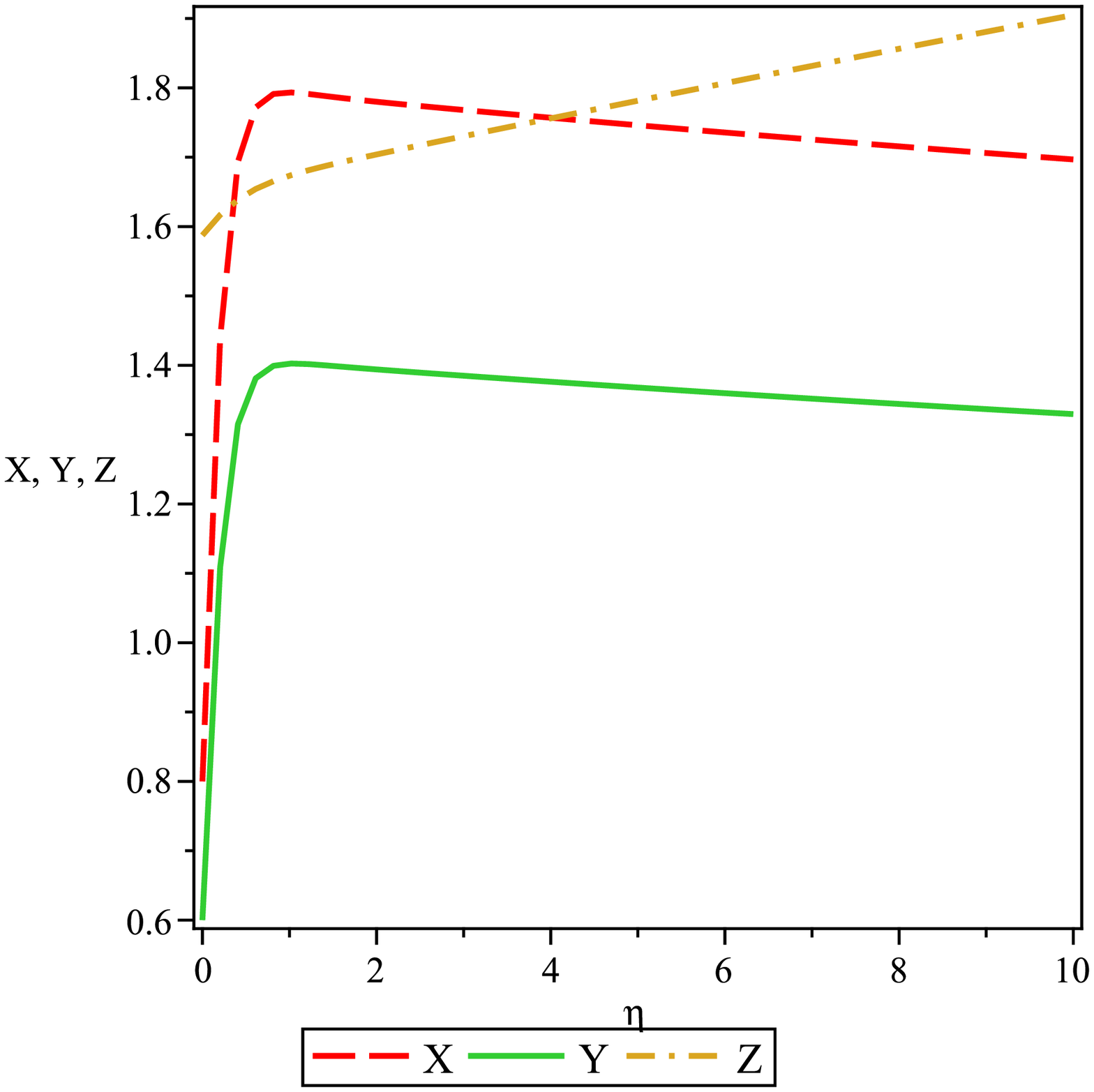}\\
~~FIG.3~~~~~~~~~~~~~~~~~~~~~~~~~~~~~~~~~~~~~~~~~~~~~~~~~~~~~~~~~~~~~~~~~~~~FIG.4\\
\caption{The phase space diagram of parameters $X(\eta), Y(\eta), Z(\eta)$ for $\gamma=-\frac{1}{2}, \omega=-0.5,~~V_0=4$.  The initial conditions chosen are $X(0) =2.2$, $Y(0) = 2.4$, $Z(0)=1.2 V_0^{\frac{2}{3}}$ (green); $X(0) = 2.3$, $Y(0) = 2.6$, $Z(0)=1.3 V_0^{\frac{2}{3}}$ (blue);
$X(0) = 2.4$, $Y(0) = 2.8$, $Z(0)=1.4 V_0^{\frac{2}{3}}$ (red); $X(0) = 2.5$, $Y(0) = 3$, $Z(0)=1.5 V_0^{\frac{2}{3}}$(brown). }

\caption{The progression of $X(\eta), Y(\eta), Z(\eta)$ for $\gamma=-\frac{1}{2},\omega=-0.5,~~V_0=2$ and initial condition is $X(0) =
0.8$, $Y(0) = 0.6$, $Z(0)=V_0^{\frac{2}{3}}$ . }
\end{figure}

{\bf{Case (c):}} Radiation: For $\gamma=4/3$, the system of
equations (20) becomes

\begin{eqnarray}
X'=-2XY~~~~~~~~~~~~~\nonumber\\
Y'=-2Y^2+\frac{Z}{2} +\frac{(7+5\omega) V_0}{3Z(3+2\omega)}\nonumber\\
Z'=2YZ~~~~~~~~~~~~~~~
\end{eqnarray}

From the above dynamical system of equations, we can seen that
there is only one feasible critical points namely $(X^{R}, 0,
\sqrt{2 \alpha V_0})$ ($``R"$ stands for Radiation) where
$\alpha=-\frac{7+5\omega}{3(3+2\omega)}$ and the feasible range
for $\omega$ becomes $-\frac{3}{2}<\omega<-\frac{7}{5}$, $X^R\neq
0$ is arbitrary real constant. In the feasible choice
$\gamma=\frac{4}{3}, ~~\omega=-1.45,~~V_0=0.5$, only critical
point becomes $(0.1, 0, 0.912871)$, eigen values are $(1.3512,
-1.3512, 0)$ depicts unstable solution. The phase space diagram of
parameters $X(\eta), Y(\eta), Z(\eta)$ and their progressions
have been drawn in figures 5 and 6 respectively. \\

\begin{figure}[!h]

\epsfxsize = 3.3 in \epsfysize = 3 in
\epsfbox{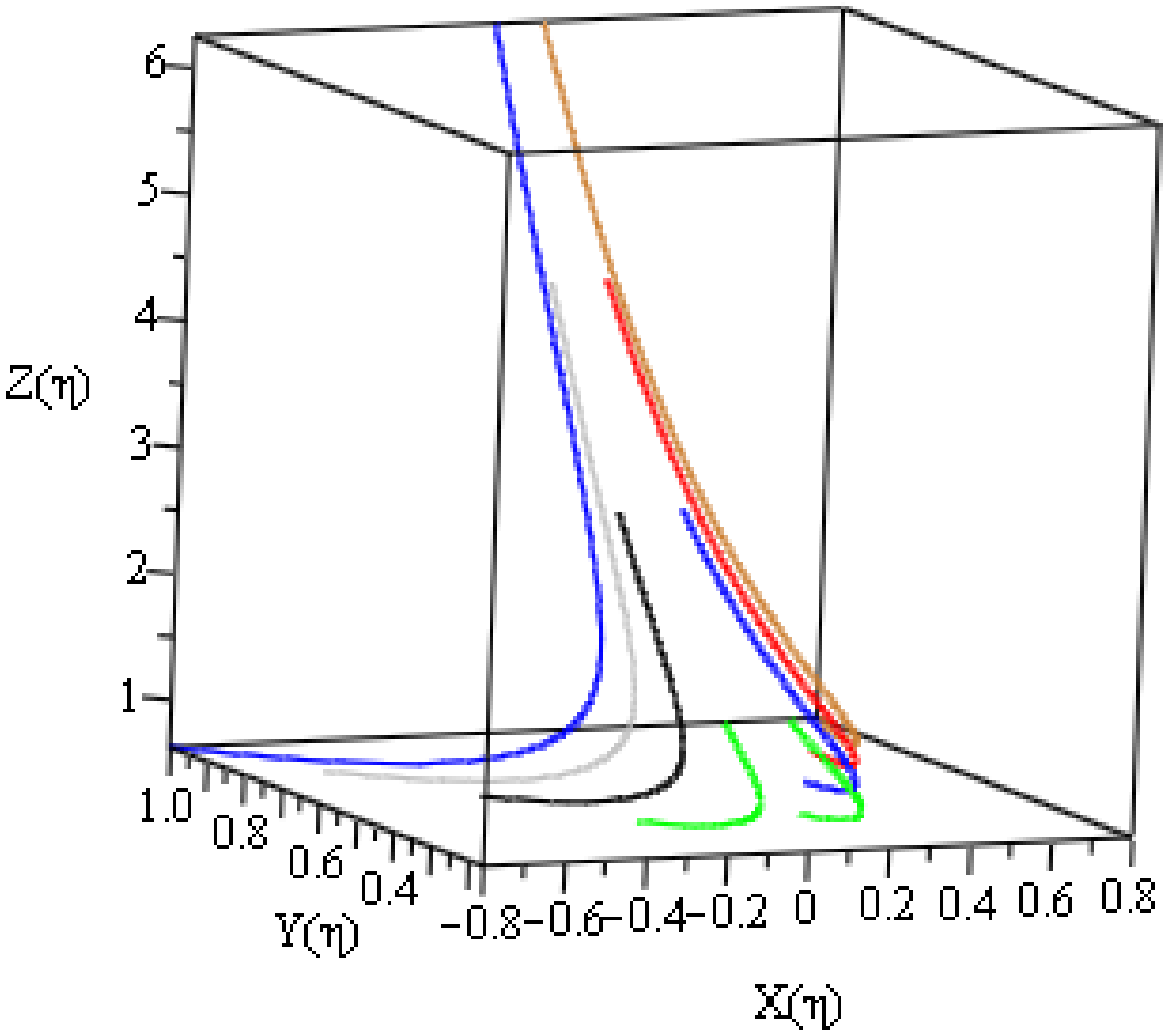}~~~\epsfxsize = 2.7 in \epsfysize = 1.7 in
\epsfbox{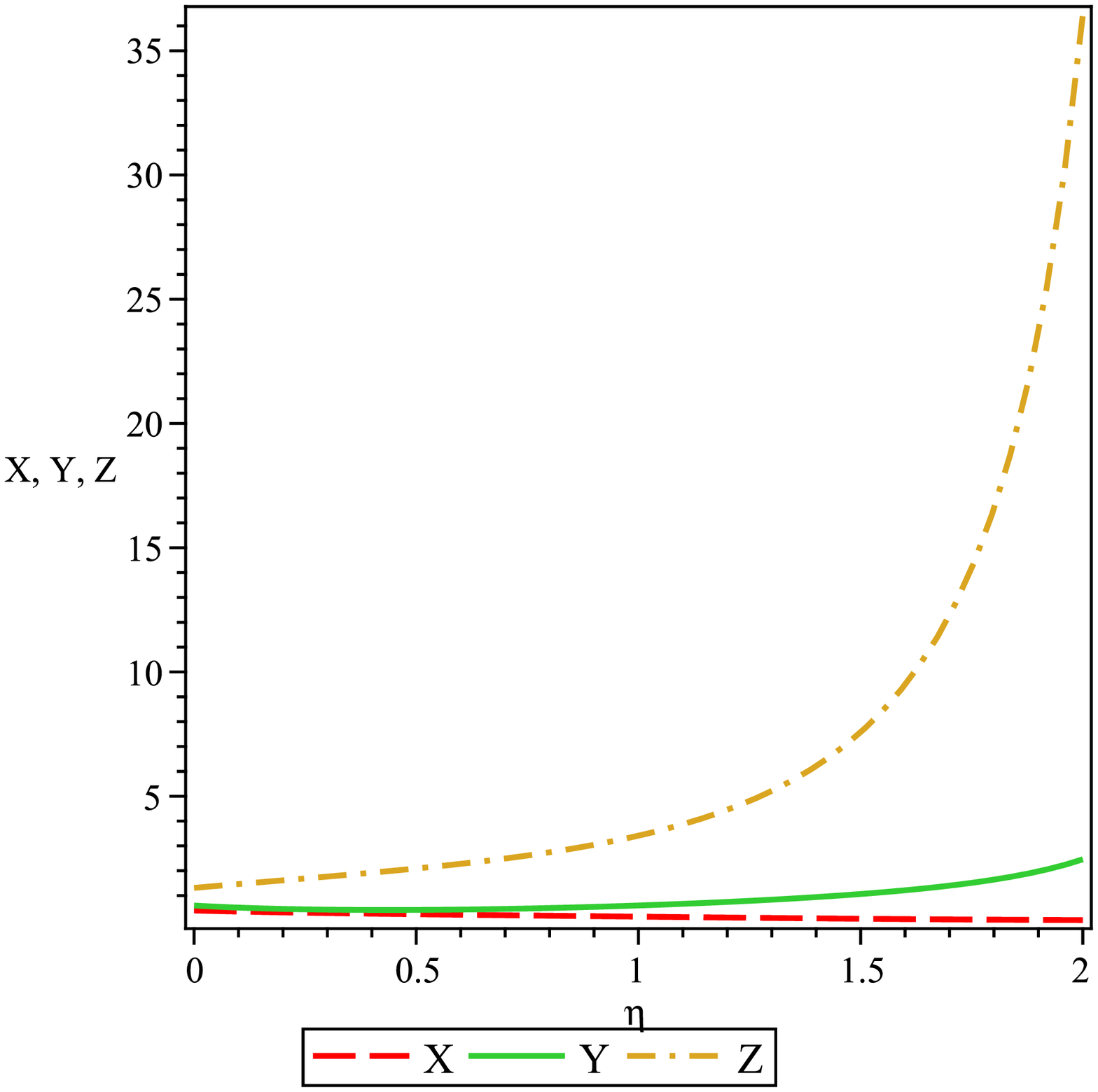}\\
~~FIG.5~~~~~~~~~~~~~~~~~~~~~~~~~~~~~~~~~~~~~~~~~~~~~~~~~~~~~~~~~~~~~~~~~~~~FIG.6\\
\caption{The phase space diagram of parameters $X(\eta), Y(\eta), Z(\eta)$ for $\gamma=\frac{4}{3}, \omega=-1.45,~~V_0=.5$.  The initial conditions chosen are $X(0) =0.2$, $Y(0) =0.4$, $Z(0)=V_0^{\frac{2}{3}}$ (green); $X(0) = 0.4$, $Y(0) = 0.6$, $Z(0)=V_0^{\frac{2}{3}}$ (blue);
$X(0) = 0.6$, $Y(0) = 0.8$, $Z(0)=V_0^{\frac{2}{3}}$ (red); $X(0) = 0.8$, $Y(0) = 1$, $Z(0)=V_0^{\frac{2}{3}}$(brown),$X(0) =-0.2$, $Y(0) =0.4$, $Z(0)=V_0^{\frac{2}{3}}$ (green); $X(0) =- 0.4$, $Y(0) = 0.6$, $Z(0)=V_0^{\frac{2}{3}}$ (blue);
$X(0) = -0.6$, $Y(0) = 0.8$, $Z(0)=V_0^{\frac{2}{3}}$ (red); $X(0) =- 0.8$, $Y(0) = 1$, $Z(0)=V_0^{\frac{2}{3}}$(brown). }

\caption{The progression of $X(\eta), Y(\eta), Z(\eta)$ for $\gamma=\frac{4}{3}, \omega=-1.45,~~V_0=1.5$ and initial condition is $X(0) =
0.4$, $Y(0) = 0.6$, $Z(0)=V_0^{\frac{2}{3}}$ . }
\end{figure}

{\bf{Case (d):}} $\Lambda$CDM: For $\gamma=0$, the system of
equations (20) becomes

\begin{eqnarray}
X'=-2XY+\frac{4V_0}{Z(3+2\omega)}+\frac{2V_0^2}{Z(3+2\omega)}\nonumber\\
Y'=-2Y^2+\frac{(5+3\omega)V_0}{Z(3+2\omega)} +\frac{(11+6\omega) V_0^2}{6Z(3+2\omega)}\nonumber\\
Z'=2XZ-2YZ~~~~~~~~~~~~
\end{eqnarray}

From the above dynamical system of equations, we can seen that
there is only one feasible critical point namely $(X^{\Lambda},
Y^{\Lambda}, Z^{\Lambda})$ ($``\Lambda"$ stands for $\Lambda$CDM)
where $X^{\Lambda}=
Y^{\Lambda}=-\frac{2\sqrt{3(1+3\omega)(4+3\omega)}}
{(1-6\omega)\sqrt{Z^{\Lambda}(3+2\omega))}}$, $Z^{\Lambda} > 0$ is
arbitrary real constant and the feasible range for $\omega$
becomes $-\frac{1}{3}<\omega<\frac{1}{6}$. In the feasible choice
$\gamma=0, ~~\omega=-0.1,~~V_0=0.2$, only critical points become
$(\pm 0.463304,\pm 0.463304, 20.2)$ and the corresponding
eigenvalues are $(\mp 1.85322, \mp 0.931517, \pm 0.00490953)$ and
so the critical points are unstable. The phase space diagram of
parameters $X(\eta), Y(\eta), Z(\eta)$
and their progressions have been drawn in figures 7 and 8 respectively. \\

\begin{figure}[!h]

\epsfxsize = 3.0 in \epsfysize = 2 in
\epsfbox{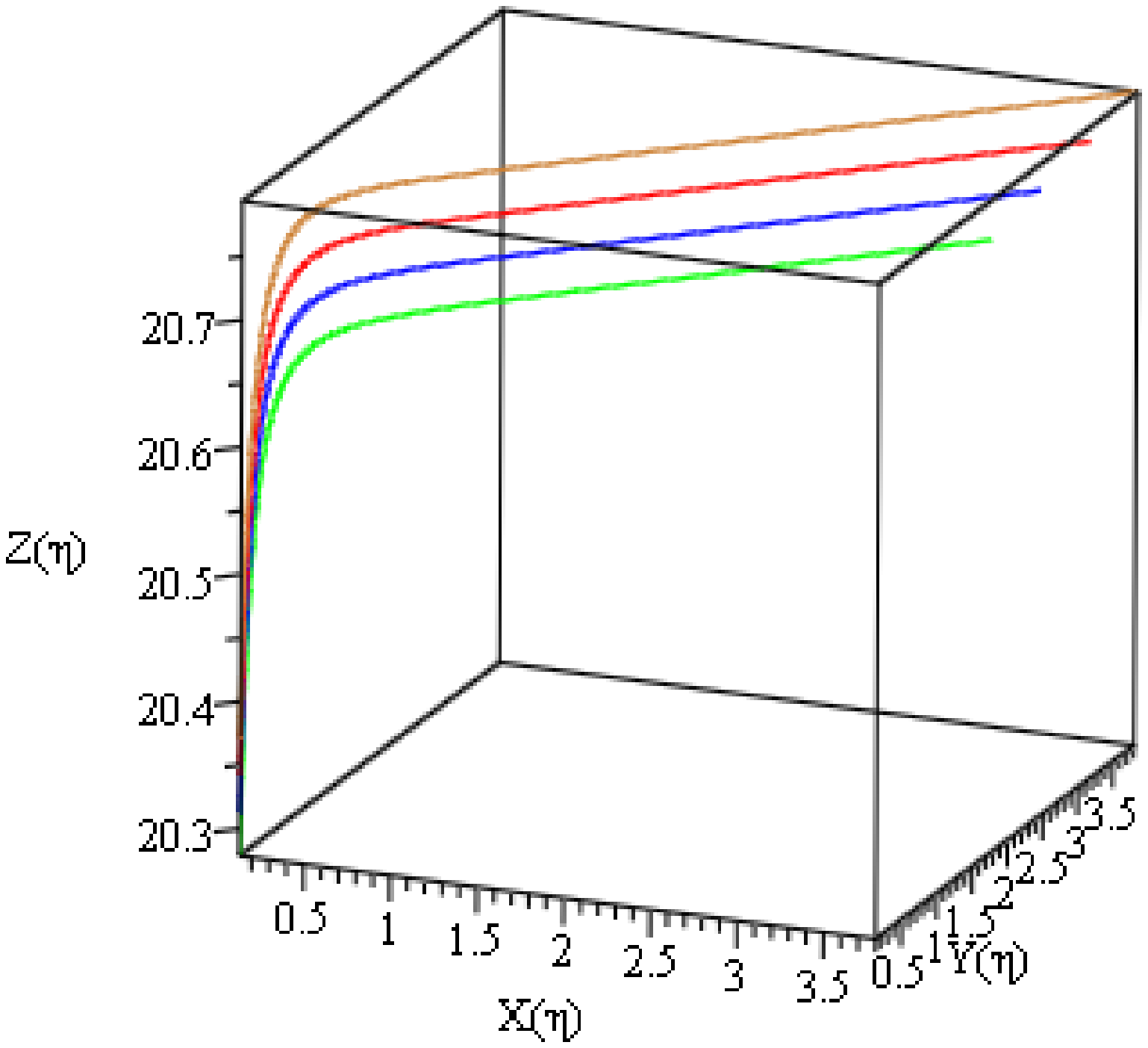}~~~\epsfxsize = 2.5 in \epsfysize = 1.7 in
\epsfbox{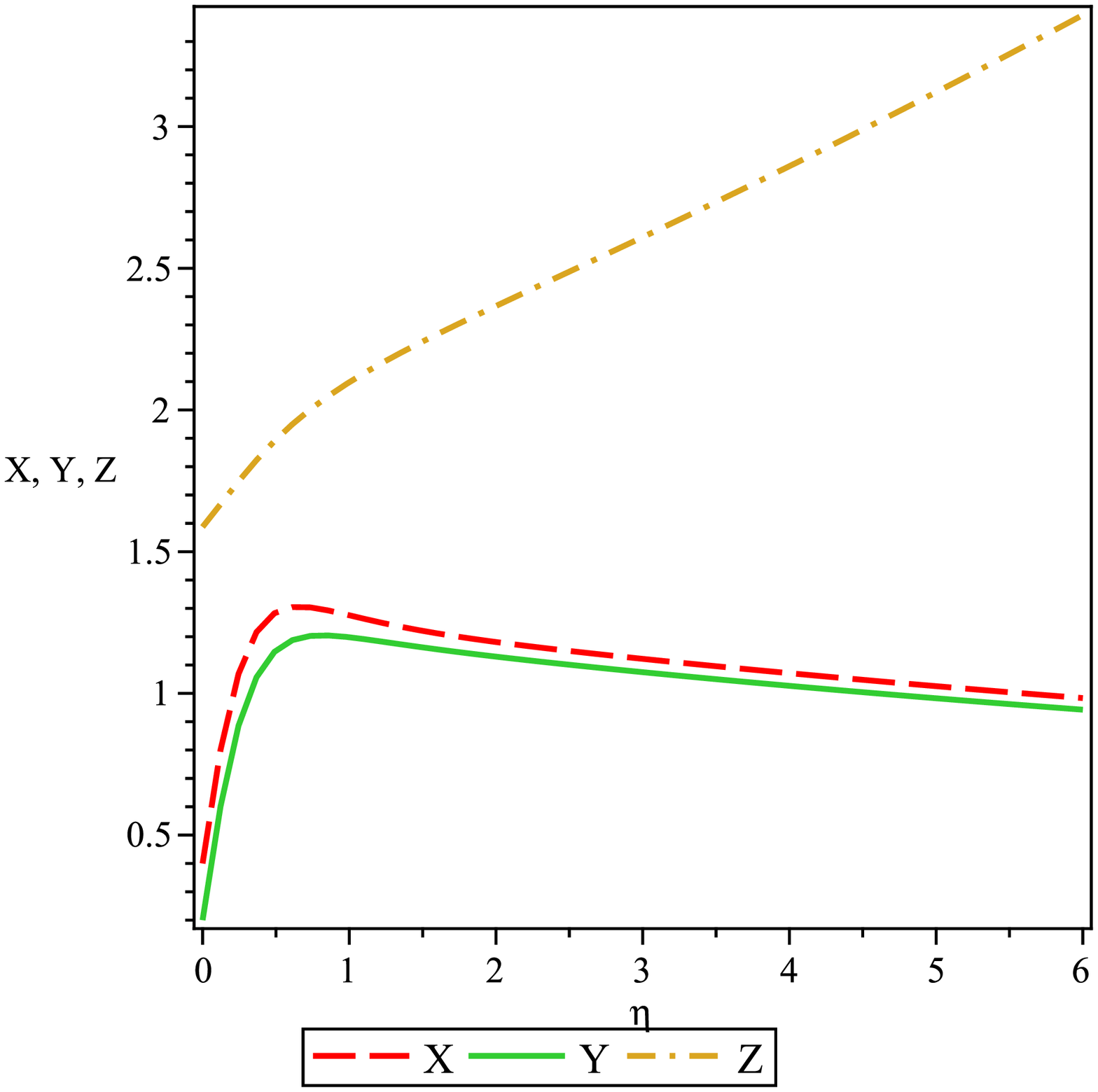}\\
~~FIG.7~~~~~~~~~~~~~~~~~~~~~~~~~~~~~~~~~~~~~~~~~~~~~~~~~~~~~~~~~~~~~~~~~~~~FIG.8\\
\caption{The phase space diagram of parameters $X(\eta), Y(\eta), Z(\eta)$ for $\gamma=0, \omega=-0.1,~~V_0=.2$.  The initial conditions chosen are $X(0) =3.2$, $Y(0) =3.2$, $Z(0)=60.5 V_0^{\frac{2}{3}}$ (green); $X(0) = 3.4$, $Y(0) = 3.4$, $Z(0)=60.6 V_0^{\frac{2}{3}}$ (blue);
$X(0) = 3.6$, $Y(0) = 3.6$, $Z(0)=60.7 V_0^{\frac{2}{3}}$ (red); $X(0) = 3.8$, $Y(0) = 3.8$, $Z(0)=60.8 V_0^{\frac{2}{3}}$(brown).}

\caption{The progression of $X(\eta), Y(\eta), Z(\eta)$ for $\gamma=0, \omega=-0.2,~~V_0=2$ and initial condition is $X(0) =
0.4$, $Y(0) = 0.2$, $Z(0)=V_0^{\frac{2}{3}}$ . }
\end{figure}

{\bf{Case (e):}} Dust: For $\gamma=1$, the system of equations
(20) becomes

\begin{eqnarray}
X'=-2XY+\frac{V_0}{Z(3+2\omega)}-\frac{Z^2}{Z(3+2\omega)V_0}\nonumber\\
Y'=-2Y^2+\frac{V_0}{Z}+\frac{Z^2(4+3\omega)}{3V_0(3+2\omega)}\nonumber\\
Z'=\frac{XZ}{2}+YZ~~~~~~~~~~~~
\end{eqnarray}

For the dust dominated universe, there are two critical points:
$\left(X^D_i, Y^D_i, Z^D_i\right)|_{i=1,2}$, where $``D"$ stands
for dust dominated universe, and given by

\begin{eqnarray}
X^D_{1,2}=\mp \frac{\left(\frac{1}{3}\right)^\frac{1}{6} \sqrt{2 (13+9\omega)V_0^\frac{1}{3}}}{(7+4\omega)^\frac{1}{6}\sqrt{3+2\omega}~(5+6\omega)^\frac{1}{3}}\nonumber\\
Y^D_{1,2}=\mp \frac{\left(\frac{1}{3}\right)^\frac{1}{6} \sqrt{ (13+9\omega)V_0^\frac{1}{3}}}{(7+4\omega)^\frac{1}{6}\sqrt{2(3+2\omega)}~(5+6\omega)^\frac{1}{3}} \nonumber\\
 Z^D_{1,2}=\frac{(21+12\omega)^\frac{1}{3}V_0^\frac{2}{3}}{(5+6\omega)^\frac{1}{3}}
\end{eqnarray}

Feasible region for existence of these critical point is
$\omega>-\frac{5}{6}$. Characteristic roots near the critical
point $\left(X^D_1, Y^D_1, Z^D_1\right)$ are two positive and one
negative real roots, and around the critical point $\left(X^D_2,
Y^D_2, Z^D_2\right)$ are two negative and one positive real roots.
In the feasible range $\gamma=1,~~\omega=-0.6,~~V_0=0.2$, critical
points are $(\pm 1.28263, \pm 0.641315, 0.733281)$, and the
corresponding eigen values are $(\mp 3.17407, \pm 0.696701,\mp
0.0878872)$ and so they are unstable. The phase space diagram of
parameters $X(\eta), Y(\eta), Z(\eta)$ and their progressions
have been drawn in figures 9 and 10 respectively. \\

\vspace{1in}
\begin{figure}[!h]

\epsfxsize = 3.3 in \epsfysize = 3 in
\epsfbox{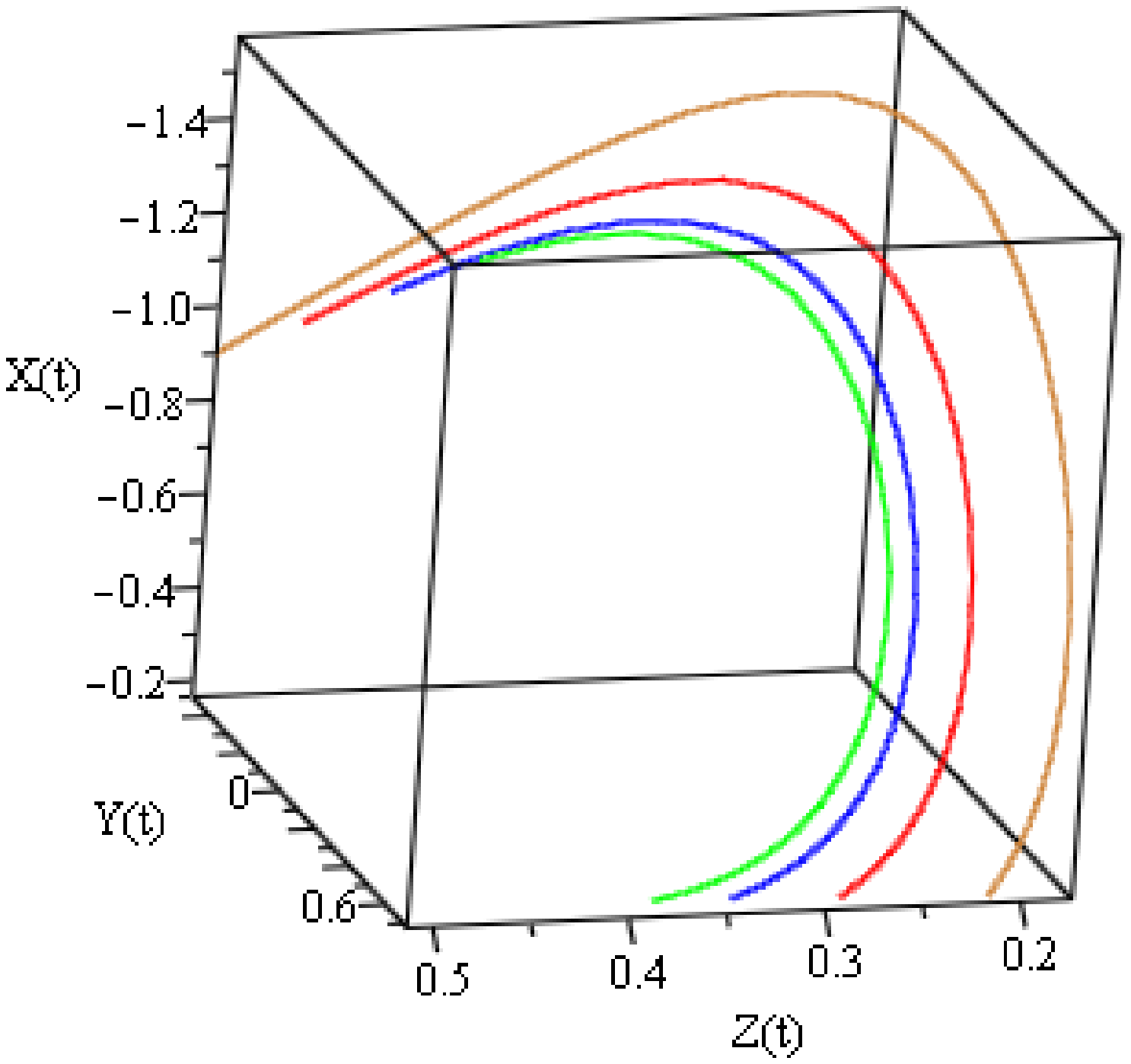}~~~\epsfxsize = 2.5 in \epsfysize = 1.7 in
\epsfbox{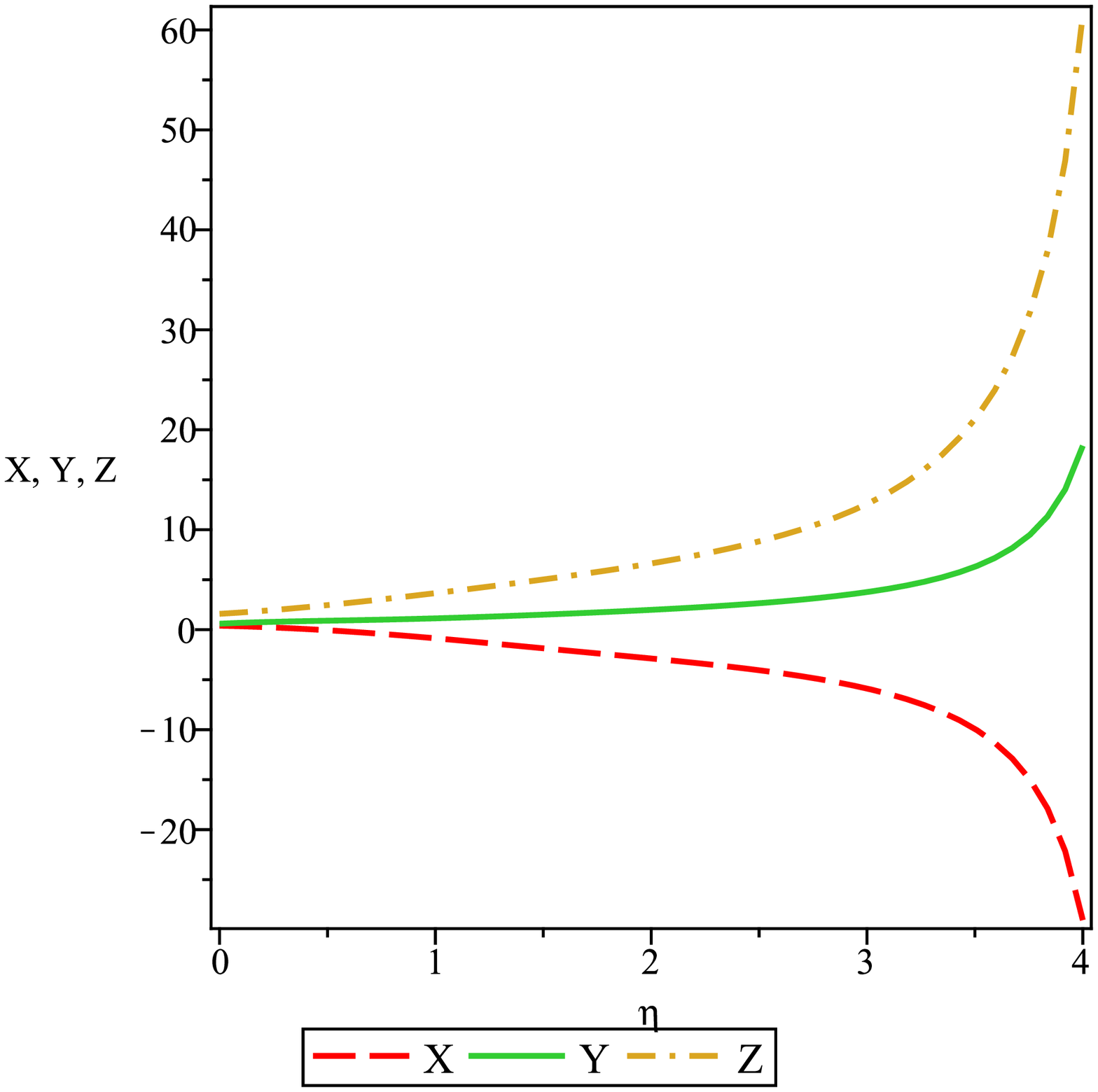}\\
~~FIG.9~~~~~~~~~~~~~~~~~~~~~~~~~~~~~~~~~~~~~~~~~~~~~~~~~~~~~~~~~~~~~~~~~~~~FIG.10\\
\caption{The phase space diagram of parameters $X(\eta), Y(\eta), Z(\eta)$ for $\gamma=1, \omega=-0.6,~~V_0=.2$.  The initial conditions chosen are $X(0) =-1.2$, $Y(0) =-0.2$, $Z(0)=1.2 V_0^{\frac{2}{3}}$ (green); $X(0) = -1.1$, $Y(0) = -0.3$, $Z(0)=1.3 V_0^{\frac{2}{3}}$ (blue);
$X(0) = -1.0$, $Y(0) = -0.4$, $Z(0)=1.4 V_0^{\frac{2}{3}}$ (red); $X(0) =-0.9$, $Y(0) = -0.5$, $Z(0)=1.5 V_0^{\frac{2}{3}}$(brown).}

\caption{The progression of $X(\eta), Y(\eta), Z(\eta)$ for $\gamma=1, \omega=-0.7,~~V_0=2$ and initial condition is $X(0) =
0.4$, $Y(0) = 0.6$, $Z(0)=V_0^{\frac{2}{3}}$ . }
\end{figure}

%
%
%

\section{\normalsize\bf{Discussions}}

In this work, we have studied the Brans-Dicke (BD) cosmology in
anisotropic models. We present three dimensional dynamical system
describing the evolution of anisotropic flat ($k=0$) model
containing perfect fluid with barotropic EoS and BD scalar field
with self-interacting potential. Choosing the power law form of
the potential function $V$ in terms of $\phi$, and defining the
three variables $X,~Y$ and $Z$, the field equations can be
transformed into the dynamical system. The critical points and the
corresponding eigen values have been found in radiation, dust,
dark energy, $\Lambda$CDM and phantom phases of the universe. The
natures and the stability around the critical points have also
been investigated. For dark energy case, we have considered
$\gamma=\frac{1}{3},~~\omega=0.5,~~V_0=0.5$, and one critical
point is $(0.679563, 1.01934 , 0.428571)$, which is a stable
attractor. Another critical point is $(-0.679563, -1.01934 ,
0.428571)$, which is unstable around that fixed point. The 3D
phase space diagram of parameters $X, Y, Z$ have been drawn in
figure 1 corresponding to the stable critical point. From figure
2, we see that $X$ and $Y$ initially increase and then decrease
and $Z$ increases in late stage of the universe. For phantom case,
we have taken $\gamma=-\frac{1}{2},~\omega=-0.5,~V_0=4$, and one
critical point is $(2.8606, 2.24761, 1.497)$, which is a stable
attractor. Another critical point is $(-2.8606, -2.24761, 1.497)$,
which is unstable around that fixed point. The 3D phase space
diagram of parameters $X, Y, Z$ have been drawn in figure 3
corresponding to the stable critical point. From figure 4, we see
that $X$ and $Y$ initially increase and then decrease and $Z$
increases in late stage of the universe. For radiation case, we
have considered $\gamma=\frac{4}{3}, ~~\omega=-1.45,~~V_0=0.5$,
and the only critical point becomes $(0.1, 0, 0.912871)$ which is
unstable. The 3D phase space diagram of parameters $X, Y, Z$ have
been drawn in figure 5 corresponding to the critical point. From
figure 6, we see that $Y$ and $Z$ increase and $X$ decreases. For
$\Lambda$CDM case, we have taken, $\gamma=0,
~~\omega=-0.1,~~V_0=0.2$, and the critical points become $(\pm
0.463304,\pm 0.463304, 20.2)$ and these are unstable. The 3D phase
space diagram of parameters $X, Y, Z$ have been drawn in figure 7
corresponding to the critical point. From figure 8, we see that
$X$ and $Y$ initially increase and then decrease and $Z$ increases
in late stage of the universe. For dust case, we have assumed
$\gamma=1,~~\omega=-0.6,~~V_0=0.2$, and the possible critical
points are $(\pm 1.28263, \pm 0.641315, 0.733281)$ which are
unstable. The 3D phase space diagram of parameters $X, Y, Z$ have
been drawn in figure 9 corresponding to the critical point. From
figure 10, we see that $Y$ and $Z$ increase and $X$ decreases. So
the anisotropic model of the universe in Brans-Dicke theory can be
stable for some cases of the fluid distribution in late stage of the
evolution.\\

{\bf Acknowledgement:}\\

One of the authors (JB) is thankful to CSIR, Govt of India for providing Junior Research Fellowship.\\

{\bf References:}\\
\\
$[1]$ C. Brans and R. H. Dicke, {\it Phys. Rev.} {\bf 124} 925
(1961).\\
$[2]$ D. A. La and P. J. Steinhardt, {\it Phys. Rev. Lett.} {\bf
62} 376 (1989).\\
$[3]$ N. Banerjee and D. Pavon, {\it Phys. Rev. D} {\bf 63} 043504 (2001).\\
$[4]$ C. Will, {\it Theory and Experiments in Gravitational
Physics} (Cambridge, Cambridge University Press) (1993).\\
$[5]$ B. K. Sahoo and L. P. Singh, {\it Modern Phys. Lett. A} {\bf 18} 2725- 2734 (2003).\\
$[6]$ K. Nordtvedt,Jr., {\it Astrophys. J} {\bf 161} 1059 (1970);
P. G. Bergmann, {\it Int. J. Phys.} {\bf 1} 25 (1968); R. V.
Wagoner, {\it Phys. Rev. D} {\bf 1} 3209 (1970); T. Damour and K.
Nordtvedt,
{\it Phys. Rev. Lett.} {\bf 70} 2217 (1993); {\it Phys. Rev. D} {\bf 48} 3436 (1993).\\
$[7]$ P. G. Bergmann, {\it Int. J. Theor. Phys.} {\bf 1} 25
(1968); R. V. Wagoner, {\it Phys. Rev. D} {\bf 1} 3209 (1970).\\
$[8]$ J. D. Barrow and K. Maeda, {\it Nucl. Phys. B} {\bf 341}
294 (1990).\\
$[9]$ C. Santos and R. Gregory, {\it Ann. Phys., (NY)} {\bf 258} 111 (1997).\\
$[10]$ O. Bertolami and P. J. Martins, {\it Phys. Rev. D} {\bf 61} 064007 (2000).\\
$[11]$ O. I. Bogoyavlensky, {\it Qualitative Theory of Dynamical
Systems in Astrophysics and Gas Dynamics}
(Springer-Verlag, Berlin, 1985).\\
$[12]$ M. Novello and C. Romero, {\it Gen. Rel. Grav.} {\bf 19}
1003 (1987).\\
$[13]$ P. Turkowski and K. Maslanka, {\it Gen. Rel. Grav.} {\bf
19} 611 (1987).\\
$[14]$ V. A. Belinskii et al, {\it Sov. Phys. JETP} {\bf 62} 195
(1986).\\
$[15]$ C. Romero and H. P. Oliveira, {CBPF-NF-}045/88, (1988); C.
Romero and A. Barros, {\it Gen. Rel. Grav.} {\bf 25} 491 (1993);
S. J. Kolitch, {\it Annals Phys.} {\bf 246} 121 (1996); D. J.
Holden and D. Wands, {\it Class. Quantum Grav.} {\bf
15} 3271 (1998).\\
$[16]$ P. Wu and H. Yu, {\it Class. Quantum Grav.} {\bf 24} 4661
(2007); H. Zhang and Z-H Zhu, {\it Phys. Rev. D} {\bf 73}  043518 (2006).\\
$[17]$ M. Jamil, {\it Int. J. Theor. Phys.} {\bf 49} 62 (2010).\\
 $[18]$ H. M. Sadjadi, {\it arxiv:} 1109.1961.\\
 $[19]$ J. Martin and M. Yamaguchi, {\it Phys. Rev. D} {\bf 77} 123508 (2008); B. Gumjudpai, T. Naskar, M. Sami and
 S. Tsujikawa, {\it JCAP } {\bf 0506} 007 (2005).\\
 $[20]$ O. Hrycyna and M. S. lowski, {\it JCAP} {\bf 04} 026 (2009).\\
 $[21]$ R-J Yang1 and X-T Gao, {\it arxiv:} 1006.4968.\\
 $[22]$ K. Xiao and J-Y Zhu,  {\it Phys. Rev. D } {\bf 83}  083501 (2011); M. Jamil, D. Momeni and M. A.
 Rashid, {\it Eur. Phys. J. C } {\bf 71} 1711  (2011); P. Wu  and S. N. Zhang, {\it JCAP} {\
 bf 0806} 007 (2008).\\
 $[23]$R. Garcia-Salcedo,  T. Gonzalez, C. Moreno, and I. Quiros, {\it arxiv:} 0905.1103.\\
$[24]$ S. Kolitch and D. Eardley, {\it Ann. Phys. (NY)} {\bf 241} 128 (1995).\\
$[25]$ S. Chakraborty, N. C. Chakraborty and U. Debnath, {\it Int.
J. Mod. Phys. D} {\bf 11} 921 (2002); {\it Int. J. Mod. Phys. A}
{\bf 18} 3315 (2003); {\it Mod. Phys. Lett. A} {\bf 18}
1549 (2003).\\
$[26]$ J. P. Mimoso and D. Wands, {\it Phys. Rev. D} {\bf 52} 5612
(1995); A. Feinstein and J. Ibanez, {\it Class. Quantum Grav.}
{\bf 10} 93 (1993).\\
$[27]$ K. S. Thorne, {\it Astrophys. J.} {\bf 148} 51 (1967).\\

\end{document}